# An Energy-Efficient Reconfigurable DTLS Cryptographic Engine for End-to-End Security in IoT Applications

Utsav Banerjee, Chiraag Juvekar, Andrew Wright, Arvind, Anantha P. Chandrakasan
Massachusetts Institute of Technology, Cambridge, MA

End-to-end security protocols, like Datagram Transport Layer Security (DTLS) [1], enable the establishment of mutually authenticated confidential channels between edge nodes and the cloud, even in the presence of untrusted and potentially malicious network infrastructure. While this makes DTLS an ideal solution for IoT, the associated computational cost makes software-only implementations prohibitively expensive for resource-constrained embedded devices. We address this challenge through three key contributions: reconfigurable cryptographic accelerators enable two orders of magnitude energy savings, a dedicated DTLS engine offloads control flow to hardware reducing program code and memory usage by ~10x, and an on-chip RISC-V core exercises the flexibility of the cryptographic accelerators to demonstrate security applications beyond DTLS.

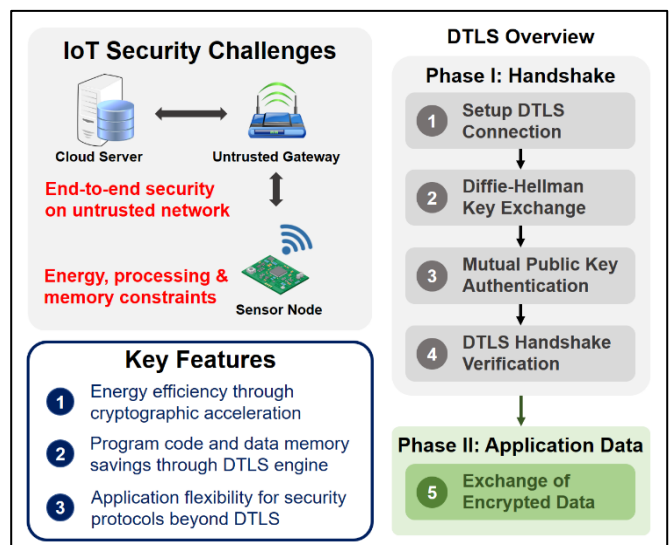

Fig. 1: End-to-end security for IoT – challenges and proposed solutions.

Fig. 1 summarizes the two major phases of the DTLS protocol: handshake and application data. The handshake phase consists of four steps. In the first step, the client (edge node) and the server agree upon protocol parameters such as the cryptographic algorithms to be used. Next, a Diffie-Hellman key exchange is performed to establish a shared secret over the untrusted channel. Following this, the client and the server authenticate each other through digital certificate verification. Finally, the two parties verify the integrity of the information exchanged in the above steps, to prevent man-in-the-middle attacks. At this point, a mutually authenticated confidential channel has been established between the client and the server, which can then be used in the second phase to exchange encrypted application data. We accelerate both phases of the DTLS protocol in hardware.

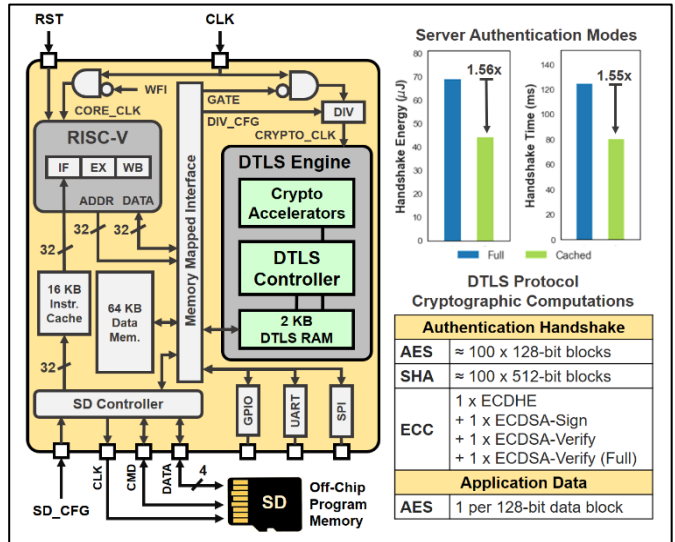

Fig. 2: System diagram with supported server authentication modes and DTLS computation details.

Fig. 2 shows the system block diagram, the DTLS modes we support, and details of the computations required to implement these modes. Our system consists of a 3-stage RISC-V processor [2] supporting the RV32I instruction set, with a 16KB instruction cache and a 64KB data memory. An SD card is used as the backing store for larger programs. A memory-mapped DTLS engine (DE), comprised of a protocol controller, a dedicated 2 KB RAM, and AES-128 GCM, SHA-256 and prime curve elliptic curve cryptography (ECC) primitives, accelerates the DTLS protocol. Sleep mode is implemented on the RISC-V, to save power, by gating its clock when cryptographic tasks are delegated to the DE. The DE uses a dedicated hardware interrupt to wake the processor on completion of these tasks. The DE is clocked by a software-controlled divider to decouple the processor operating frequency from the long critical paths in the ECC accelerator. In addition to full verification of the server certificate in step three of the handshake phase, the DE also supports caching of server certificate information to speed up future handshakes. This cached mode reduces an ECDSA-Verify operation to gain 1.56x savings in handshake energy.

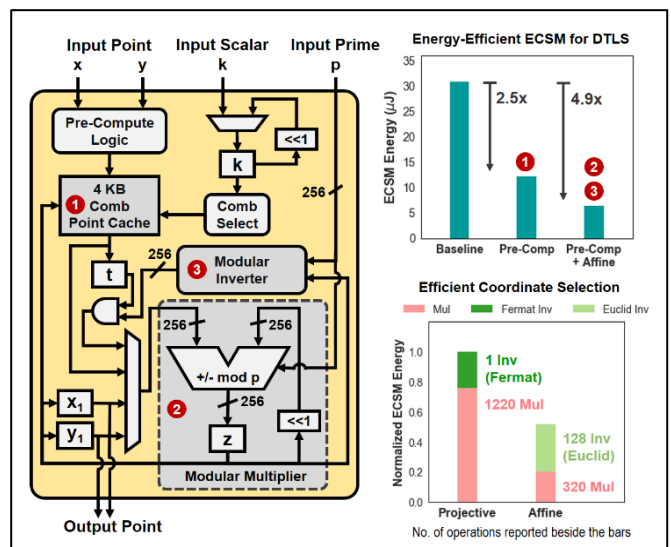

Fig. 3: Proposed prime field elliptic curve cryptography accelerator and energy savings.

Even in the cached mode, ECC computations, such as ECDHE and ECDSA, account for over 99% of the DTLS handshake energy. Fig. 3 describes an energy-efficient ECC accelerator that reduces this overhead. A pre-computation-based comb algorithm [3] is used for elliptic curve scalar multiplication (ECSM), and a 4KB cache can store pre-computed comb data for up to six points, including generator points and public keys, thus reducing ECSM energy by 2.5x compared to a baseline implementation. A 256-bit wide interleaved reduction-based modular multiplier is implemented to support all Weierstrass and Montgomery curves over prime fields up to 256 bits, with higher bits of the data-path gated when working with smaller primes. The use of interleaved reduction allows us to handle arbitrary primes without any special structure, enabling support for NIST, SEC and ANSI curves. Resource-constrained ECC implementations [4,5] typically use projective coordinates to avoid modular inversion in the ECSM inner loop, at the cost of extra multiplications and a final expensive Fermat inversion. This work implements a dedicated 31k-gate modular inverter, allowing the use of affine coordinates, which saves 1.93x in ECSM energy by trading off the extra multiplications for cheaper Euclid inversions. Furthermore, a zero-less signed digit representation [3] of the scalar k is used to prevent simple power analysis side-channel attacks on the ECSM.

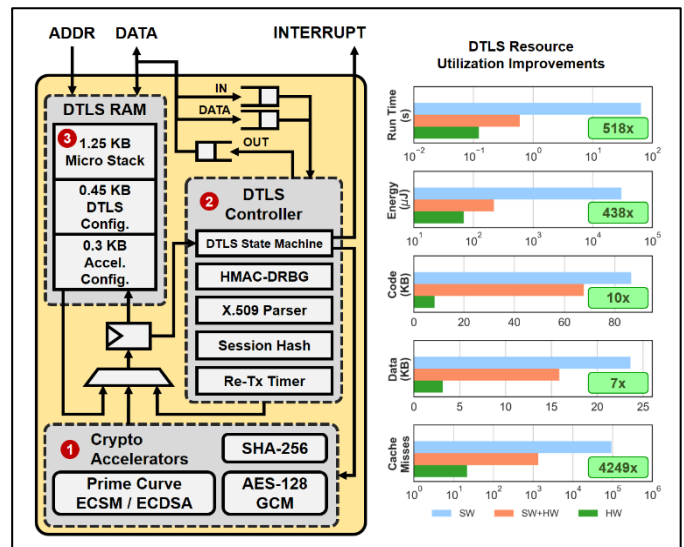

Fig. 4: Architecture of the DTLS Engine with resource utilization improvements over software-only and software-plus-hardware implementations.

Fig. 4 shows the detailed architecture of the DTLS engine and a comparison of resource utilization in three scenarios: DTLS fully implemented as RISC-V software (SW), the cryptographic kernels accelerated in hardware and only the DTLS controller implemented in software (SW+HW), and DTLS fully implemented in hardware (HW). Three blocks in the DE, that result in resource utilization improvement, are highlighted in Fig. 4. The use of cryptographic accelerators alone results in over 2 orders of magnitude improvement in run time and energy efficiency (SW vs. SW+HW). Similarly, the elimination of ECC code reduces instruction cache thrashing by 70x. Next, the DTLS controller implements a micro-coded state machine for packet framing, computation of the session transcript, parsing and validation of X.509 digital certificates and HMAC-DRBG-based pseudo-random number generation. This reduces code size by ~60KB and instruction cache misses by 60x (SW+HW vs. HW). Finally, the DTLS RAM implements a micro stack for storing temporary variables computed during the DTLS handshake. This 1.25KB micro stack results in 13KB reduction in data memory usage on the RISC-V (SW+HW vs. HW).

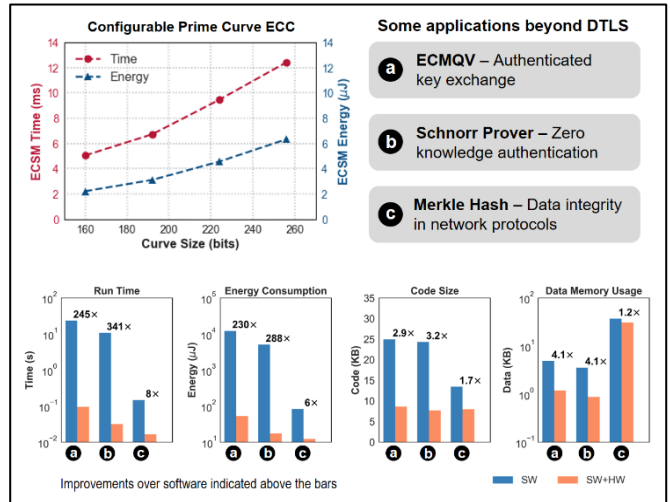

Fig. 5: Flexibility of the DTLS Engine with reconfigurable ECC and application benchmarks.

Fig. 5 demonstrates the reconfigurability of the DE. The ECC primitive in the DE can accelerate all prime curves up to 256 bits. ECSM energies and run times as a function of prime bit-width, using the secp160r1, secp192r1, secp224r1 and secp256r1 curves, are shown in Fig. 5. Security applications beyond DTLS can be implemented on the RISC-V, using the cryptographic accelerators in standalone mode. We illustrate this flexibility using three benchmark applications: (a) ECMQV, an alternative to ECDHE+ECDSA-based authenticated key exchange, (b) Schnorr Prover, an interactive zero-knowledge prover of identity, and (c) Merkle Hashing, used to ensure data integrity in peer-to-peer network protocols. The reduction in resource utilization for all three applications is shown in Fig. 5. The ECC-based applications experience over 200x increase in energy efficiency, while Merkle hashing sees 6x energy savings.

| Specifications | WISTP'11 [4] [a] | CHES'15 [5] [a] | VLSIC'17 [6] | This work |
|---|---|---|---|---|
| Technology | 350 nm | 130 nm | 40 nm | 65 nm |
| Supply voltage (V) | 3.3 | 1.2 | 0.7 | 0.8 |
| Frequency (MHz) | 0.847 | 1 | 28.8 | 16 |
| App. Processor | – | – | ARM Cortex M0 | RISC-V RV32I |
| Cryptographic Accelerator | | | | |
| Logic gates | 12.8k | 32.6k | – | 149k |
| SRAM | 0.25 KB | 0.28 KB | 8 KB | 6.75 KB |
| Hardware ECSM support | Only NIST P-192 | Only Curve25519 | – [b] | All prime curves up to 256 bits |
| Base point ECSM energy ($\mu$J) | 1423.6 (192 bit) | 56.8 (255 bit) | – [b] | 3.11 (192 bit) 6.34 (256 bit) |
| AES energy (nJ) | 8558.04 | 521.01 [c] | 7.05 | 6.21 |
| SHA energy (nJ) | 6876.3 [d] | – | 48.7 [d] | 24.3 [d] |
| DTLS in hardware | No | No | No | Yes |
| DTLS energy | – | – | – | 44.08 $\mu$J (Handshake) 0.89 nJ/B (App. Data) |

[a] Post-synthesis data reported for [4] and [5]
[b] [6] implements only modular multiplication in binary fields in hardware
[c] [5] implements Salsa20 instead of AES for encryption
[d] [4] implements SHA-1; [6] implements SHA-3; This work implements SHA-2

Fig. 6: Comparison with integrated cryptographic accelerators for embedded systems.

Fig. 6 compares this work with embedded systems that integrate multiple cryptographic accelerators. This work implements a flexible ECC accelerator which supports arbitrary primes up to 256 bits, in contrast with [4] and [5] which only support fixed 192 and 255-bit curves respectively. [6] only supports binary field modular multiplication in hardware. Our ECC accelerator is 458x and 9x more energy-efficient than [4] and [5] respectively at comparable security levels. In addition to the resource savings enabled by the individual cryptographic accelerators, offloading DTLS control flow to the DE realizes a further 3x reduction in energy and 5x reduction in run time.

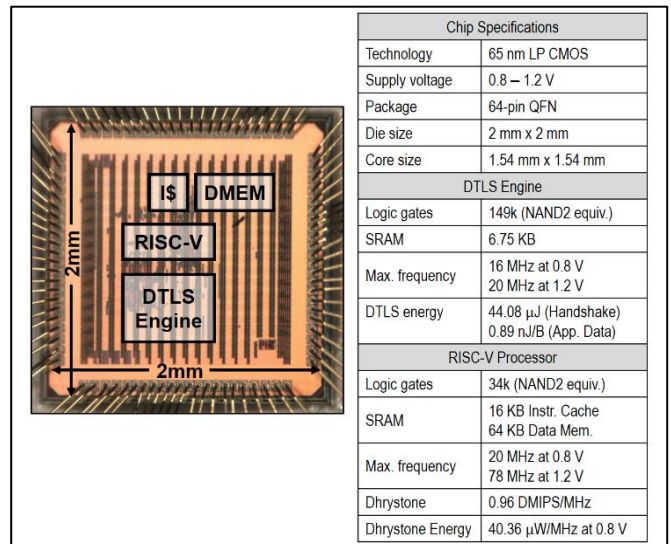

Fig. 7: Chip micrograph and performance summary.

The chip was fabricated in a 65nm LP CMOS process, occupies 2.4mm$^2$ active area, and supports voltage scaling from 1.2V down to 0.8V. All measurements for the RISC-V and the DE are reported at 16MHz and 0.8V. The RISC-V processor occupies 34k NAND Gate Equivalents (GE), and achieves 0.96 DMIPS/MHz when consuming 40.36 µW/MHz. The DTLS engine occupies 149k GE and uses 6.75KB of SRAM. The DE consumes 44.08 µJ per DTLS handshake, and 0.89 nJ per byte of application data. Therefore, through the design of reconfigurable energy-efficient cryptographic accelerators and a dedicated protocol controller, this work makes DTLS a practical solution for implementing end-to-end security on resource-constrained IoT devices.

*Acknowledgements:*

The authors would like to thank the Qualcomm Innovation Fellowship and Texas Instruments for funding this work, and the TSMC University Shuttle Program for chip fabrication support.